\documentclass[12pt]{article}
\usepackage{amssymb,amsmath,epsfig, float}
\allowdisplaybreaks
\begin{document}

\title{\bf Impact of $f(\Re,\mathcal{T}^{2})$ Theory on Stable Finch-Skea Gravastar
Model}
\author{M. Sharif$^1$ \thanks {msharif.math@pu.edu.pk} and Saba Naz$^2$
\thanks{sabanaz1.math@gmail.com}\\
$^1$ Department of Mathematics and Statistics, The University of Lahore,\\
1-KM Defence Road Lahore, Pakistan.\\
$^2$ Department of Mathematics, University of the Punjab,\\
Quaid-e-Azam Campus, Lahore-54590, Pakistan.}

\date{}

\maketitle

\begin{abstract}
This paper examines the structure of a gravitationally vacuum star
(also known as gravastar) in the background of
${f}(\Re,\mathcal{T}^{2})$ theory. This hypothetical object can be
treated as a substitute of a black hole, with three regions: (i) the
internal region, (ii) the intrinsic shell and (iii) the outer
region. We examine these geometries using Finch-Skea metric for
radial metric component along with particular
$f(\Re,\mathcal{T}^{2})$ model. We determine singularity-free
solution for both the inner as well as thin-shell domains. The
smooth matching of the interior region with external Schwarzschild
spacetime is obtained through Israel matching constraints. Finally,
we study various characteristics of gravastar domains including
equation of state parameter, proper length, entropy, energy as well
as surface redshift. It is found that the compact gravastar
structure is a viable alternate to the black hole in the perspective
of this gravity.
\end{abstract}

\section{Introduction}

Cosmic systems incorporating both small as well as large-scale
celestial bodies have an impact on the evolution of the universe and
provide a platform for astrophysical studies. Different theories
have been formulated to comprehend the mechanism and structure of
these stellar objects. Einstein formulated general relativity (GR)
to investigate the dynamical relationship between space, time,
matter, as well as curvature. According to the proposal of Edwin
Hubble, all galaxies in the cosmos are moving rapidly farther from
us resulting in accelerated cosmic expansion. Various observational
evidences have revealed the quick expanding behavior of the cosmos
\cite{sc}. The cryptic force possessing huge negative pressure
referred to as dark energy is assumed to be responsible for this
cosmic behavior.

Acceptable cosmic models have been developed by numerous researchers
to describe the universe origin, evolution as well as transition
through many cosmic epochs. In accordance of big-bang theory, cosmic
matter and energy were occupied in a tiny region, also described as
singularity with an endless temperature and energy density. Although
the big-bang theory is widely accepted, many more fascinating
proposals have been put forward by experts to explain the origin and
evolution of the cosmos. The universe evolution can also be
explained by bounce cycles (continuous expanding as well as
contracting behavior) with neither a beginning nor an end and is
named as the bounce theory.

Katirci and Kavuk \cite{7a} introduced the revolutionary notion of
bounce theory to formulate a generalization of GR. Through the
contraction of energy-momentum tensor (EMT), they established a
particular relation between matter and gravity dubbed as
\textsf{energy-momentum squared gravity (EMSG)} or
$f(\Re,\mathcal{T}^{2})$ theory with $\mathcal{T}^{2}=T^{\alpha
\vartheta}T_{\alpha \vartheta}$. This theory is found to be a
promising framework for avoiding the big-bang singularity since it
has a minimum scale factor and maximum finite density \cite{36}. In
this theory, the role of cosmological constant is to resolve
singularity by providing a repulsive force. This theory follows the
true sequence of cosmic epochs and effectively explains the cosmic
behavior. The field equations in this theory incorporate squared and
product components of matter variables, which significantly
contribute to the investigation of many cosmological scenarios.

Numerous researchers have contributed extensive work regarding this
modified theory. Roshan and Shojai \cite{36} solved EMSG field
equations with homogeneous perfect matter configuration to compute
exact solutions and determined the possibility of bouncing cosmos at
early times. The range of exact solutions for isotropic expanding
universe in relation to the early and late-time cosmic evolution was
derived by Board and Barrow \cite{37}. The correlation between mass
as well as radius of a neutron star was presented in \cite{38}. They
found that smaller or larger size of these stars is controlled by
the core pressure along with the model parameter of this gravity.
Moraes and Sahoo \cite{17} investigated wormhole solutions with
non-exotic matter whereas Akarsu et al. \cite{18} studied probable
neutron star constraints in the same framework. Bahamonde et al.
\cite{41} examined the role of minimal/non-minimal coupled models
and concluded that these models well describe the cosmic history and
accelerated expansion.

Barbar et al. \cite{21} studied feasible constraints of the bouncing
cosmos in this theory and found non-singular universe in the early
times. Ranjit et al. \cite{22} used cosmic chronometer as well as
Supernovae type-Ia data to study cosmological implications of the
solutions for matter density of the proposed EMSG model. Singh et
al. \cite{23} used quark matter to explore the structure of
celestial bodies. Sharif and Gul \cite{42} discussed various
scenarios such as gravitational collapse, Noether symmetry and
stability of the Einstein universe in the realm of this theory. We
have explored different attributes of gravastar solutions in the
presence/absence of electromagnetic field \cite{44}. Recently,
Sharif and his collaborators computed solutions through minimal
geometric deformation \cite{44*} and studied the complexity of
charged static sphere as well as uncharged static cylinder in this
gravity \cite{44**}.

The fundamental elements of the galaxies arrayed in a cosmic web are
stars which are made of hydrogen and helium. The equilibrium of
stars against inward-directed gravitational pull is maintained by
burning their fuel. When a star burns all of its fuel in the core,
then outer pressure vanishes. Consequently, new compact objects are
formed as a final outcome of the gravitational collapse of the star.
One of the highly dense objects is an entirely collapsed body
regarded as a black hole. In black hole geometry, the event horizon
surrounds singularity (known as the place of no return). The
pioneering work on stellar structure referred to as gravastar was
presented by Mazur and Motolla \cite{13m} to study \emph{singularity
as well as event horizon} issues. The appealing attribute of this
novel compact structure is its \emph{singularity-free} nature. They
used the de Sitter (dS) interior to prevent the singularity whereas
a very thin layer of baryonic matter disintegrates the outer and
inner regions. An equation of state (EoS) governs the
characteristics of each domain.

Currently, there is no observational evidence in support of
hypothetical gravastar, however, numerous studies indicate the
existence as well as detection of gravastar structure in the near
future. The criteria for detection of gravastars can be carried out
by studying gravastar shadows \cite{d1}. To detect gravastar
geometries, gravitational lensing (measurement of the overall
changing luminosity of companion or nearby stars) was proposed by
\cite{d2}. In contrast to black holes, we may find a maximum
luminosity effect in gravastar structure with a mass equal to that
of a black hole. The observation of GW150914 through interferometric
LIGO detectors showed ringdown signals hinting towards the
possibility of an entity having no event horizon \cite{d3}.
Furthermore, a current investigation through \texttt{First M87 Event
Horizon Telescope (EHT)} was made and suggested that captured
shadows might be a part of the gravastar geometry \cite{d5}.

Many astrophysicists are paying close attention to gravastars in an
effort to understand more about their structural properties. For
particular EoS parameters, Visser and Wiltshire \cite{5} found
stable configurations by examining the impact of radial
perturbations on gravastar stability. An extension to this work was
given by presenting the vacuum energy bounds for inner and outer
sectors of gravastar structure \cite{6}. To compute solutions for an
extended range of radii as well as masses, Bili\'{c} et al. \cite{7}
used the dS geometry in the internal domain in place of Born-Infeld
phantom spacetime. To study various attributes against shell
thickness, an investigation of gravastar solutions incorporating the
Kuchowicz metric ansatz was presented by \cite{34}. Ghosh and his
collaborators \cite{36*} determined gravastar solutions by employing
the Karmarkar condition for both the interior as well as shell
region.

Das et al. \cite{29} investigated various characteristics of
intermediate shell of gravastar in $f(\Re,\mathcal{T})$ gravity and
explored some of its features graphically. Shamir and Ahmad
\cite{29tt} inspected gravastar geometry in the context of
$f(\mathcal{G},\mathcal{T})$ theory (\emph{where $\mathcal{G}$ is
the Gauss-Bonnet invariant}). Sharif and Waseem \cite{28}
investigated gravastar in $f(\Re,{\mathcal{T}})$ theory using
conformal motion and found the object having no singularity. The
same authors \cite{35} discussed different features of gravastar
geometry through Kuchowicz ansatz. In recent years, several studies
\cite{702}-\cite{714} have been devoted to investigate the effects
of modified gravity on compact objects, providing important insights
into their structural properties and evolution.

The Finch-Skea metric was initially introduced as a correction to
the Dourah and Ray \cite{21} metric, which was found unsuitable for
describing compact objects. Finch and Skea \cite{22} made
modifications to this metric to accommodate relativistic stellar
models. However, they further extended the Finch-Skea metric in four
dimensions to include anisotropic stellar models
\cite{22a}-\cite{22e}. The significance of the Finch-Skea metric
lies in its utility for describing compact stellar configurations in
spacetime dimensions equal to or greater than four \cite{22f}. These
metric potentials have attracted considerable attention from
researchers due to their non-singular and viable properties. As a
result, they have been extensively employed to study compact stellar
structures with diverse matter distributions. For instance, Bhar
\cite{22g} utilized the Finch-Skea metric potentials and the
chaplygin gas EoS to analyze the physical properties of strange
stars.

In this paper, we employ the Finch-Skea metric to study gravastar
geometry in EMSG theory. We examine the graphical behavior of
various attributes of gravastar for a specific model corresponding
to the intrinsic shell. The paper is arranged as follows. Section
\textbf{2} presents basic formalism of $f(\Re,\mathcal{T}^2)$ field
equations with Finch-Skea metric component. We discuss three regions
of the gravastar geometry with respective EoS in section \textbf{3}.
In section \textbf{4}, we analyze the obtained solutions graphically
corresponding to a specific value of the model parameter. Section
\textbf{5} presents stability of gravastar structure. The summary of
the obtained findings is provided in the last section.

\section{Energy-Momentum Squared Theory}

In this section, we use perfect matter distribution to develop the
field equations of $f(\Re,\mathcal{T}^{2})$ theory. The action of
this theory is given by \cite{7a}
\begin{equation}\label{1}
\textsl{I}=\frac{1}{2\kappa^2}\int d^4x\left[f\left(\Re,T^{\alpha
\vartheta}T_{\alpha \vartheta}
\right)+\mathfrak{L}_{m}\right]{\sqrt{-g}},
\end{equation}
where $\mathfrak{L}_{m}$ is the matter Lagrangian density,
$\kappa^2=1$ denotes the coupling constant and $g$ represents
determinant of the metric tensor. The corresponding field equations
are
\begin{equation}\label{2}
\Re_{\alpha\vartheta}f_{\Re}-\frac{1}{2}g_{\alpha\vartheta}f+g_{\alpha\vartheta}\Box
f_{\Re}-\nabla _{\alpha}\nabla_{\vartheta}f_{\Re}
=T_{\alpha\vartheta}-\circleddash_{\alpha\vartheta}f_{\mathcal{T}^{2}},
\end{equation}
where $\Box={\nabla_{\alpha} \nabla^{\alpha}}$,$~f\equiv
f(\Re,\mathcal{T}^{2}) , ~f_{\mathcal{T}^{2}}=\frac{\partial
f}{\partial \mathcal{T}^2}, f_{\Re}=\frac{\partial f}{\partial
\Re}$, and $\circleddash_{\alpha\vartheta}$ is
\begin{eqnarray}\label{3}
\circleddash_{\alpha\vartheta}
=-4\frac{\partial^{2}\mathfrak{L}_{m}}{\partial g^{\alpha\vartheta}
\partial g^{\beta\eta}}T^{\beta\eta}-2\mathfrak{L}_{m}\left(T_{\alpha\vartheta}-\frac{1}{2}g
_{\alpha \vartheta }T\right)-TT_{\alpha
\vartheta}+2T_{\alpha}^{\beta}T_{\vartheta\beta}.
\end{eqnarray}
The EMT describes the relationship between the distribution of
matter and energy. Its non-zero components yield physical variables
which determine different dynamical characteristics of
self-gravitating systems. The non-conserved EMT in this gravity
suggests the occurrence of an extra force which governs the
non-geodesic particle motion as follows
\begin{equation}\label{4}
\nabla^{\alpha}T_{\alpha\vartheta}=\frac{1}{2} \left[-f_{
\mathcal{T}^{2}}g_{\alpha\vartheta}{\nabla^{\alpha}}{\mathcal{T}^{2}}+2{\nabla^{\alpha}}(f_{
\mathcal{T}^{2}}{\circleddash_{\alpha\vartheta}})\right].
\end{equation}
We consider isotropic matter as
\begin{equation}\label{5}
T_{\alpha\vartheta}=\mathcal{U}_{\alpha}
\mathcal{U}_{\vartheta}(P+\varrho)-P g_{\alpha\vartheta},
\end{equation}
where $\mathcal{U}_{\vartheta}$ symbolizes four-velocity, $\varrho$
and $P$ describe density and pressure of fluid configuration,
respectively. Employing Eqs.(\ref{3}) and (\ref{5}) with
$\mathfrak{L}_{m}=P$, we have
\begin{eqnarray}\nonumber
\circleddash_{\alpha\vartheta} = -\mathcal{U}_{\alpha}\mathcal{U}
_{\vartheta}\left(3P^2+\varrho^2+4 P\varrho\right).
\end{eqnarray}

The field equations (\ref2) involve product as well as squared
matter components, and hence become difficult to solve. Thus, we use
a \textsl{minimal/non-minimal} model of this gravity. The field
equations relating to the non-minimal model becomes complicated,
which makes their solutions very challenging. Consequently, we
employ a minimally coupled model of the form \cite{7a}
\begin{equation}\label{6}
f(\Re,\mathcal{T}^{2})= \Re+\chi \mathcal{T}^{2},
\end{equation}
where $\chi$ denotes a model parameter and
$\mathcal{T}^{2}=\varrho^{2}+3P^{2}$. This model represents the
cosmic evolution and expanding behavior. The inclusion of the
component $\mathcal{T}^{2}$ results in the extension of $f(\Re)$ as
well as $f(\Re,T)$ theories beyond GR. Different aspects of several
celestial objects have been studied using the functional form
(\ref6). It significantly resolves issues related to the universe
\cite{ab} as well as describes three cosmic epochs, known as
\texttt{dS-dominated, radiation-dominated as well as
matter-dominated} eras. Employing this relation of
$f(\Re,\mathcal{T}^{2})$ in Eq.(\ref{2}), we obtain
\begin{equation}\label{9}
\mathbb{G}_{\alpha\vartheta}= T_{\alpha\vartheta}+\frac{1}{2}\chi
g_{\alpha\vartheta} {\mathcal{T}}^{2}-\chi f_{
{\mathcal{T}}^{2}}{\circleddash_{\alpha\vartheta}},
\end{equation}
where Einstein tensor is denoted by $\mathbb{G}_{\alpha\vartheta}$.

We take a static sphere to investigate the inner geometry of
gravastar structure as
\begin{equation}\label{9}
ds^{2}=e^{\sigma(r)}dt^{2}-e^{\varpi(r)}dr^{2}-r^{2}(d\theta^{2}+\sin^{2}\theta
d\varphi^{2}).
\end{equation}
The respective field equations are
\begin{eqnarray}\label{10}
&&-e^{-\varpi}r^{-2}+ \frac{1}{r^{2}}+
\frac{\varpi^{'}r^{-1}}{e^{\varpi}}=\varrho + \frac{1}{2}\chi
\varrho^{2} + \frac{3}{2} \chi P^{2} +4 \chi \varrho P,
\\\label{11}
&&e^{-\varpi}\left(\frac{1}{r^{2}}+\frac{\sigma^{'}}{r}\right) -
\frac{1}{r^{2}}=P - \frac{\chi}{2}\varrho^{2} -\frac{3}{2} \chi
P^{2},\\\label{12}
&&e^{-\varpi}\left(\frac{\sigma^{'2}}{4}+\frac{\sigma^{'}}
{2r}-\frac{\varpi^{'}}{2r}+\frac{\sigma^{''}}{2}
-\frac{\sigma^{'}\varpi^{'}}{4}\right)=P
-\frac{\chi}{2}\varrho^{2}-\frac{3}{2}\chi P^{2},
\end{eqnarray}
where the radial derivative is denoted by prime. The radial metric
component $e^{\varpi(r)}$ is considered to be of the form proposed
by Finch-Skea and is given as \cite{7b}
\begin{equation}\label{14}
e^{\varpi(r)}=1+\aleph r^{2},
\end{equation}
where $\aleph$ is an arbitrary constant. Plugging Eq.(\ref{14}) in
(\ref{10})-(\ref{12}), we have
\begin{eqnarray}\label{15}
\aleph+2\mathfrak{F}&=&\frac{\aleph}{\mathfrak{F}}\Big[\varrho+\frac
{\chi(\varrho^{2} + {3}P^{2}+\varrho P)}{2}\Big],
\\\label{16}
\aleph-\frac{\sigma'}{r}&=&\frac{\aleph}{\mathfrak{F}}(P -
\frac{\chi}{2}\varrho^{2} -\frac{3}{2} \chi P^{2}),
\\\label{17}
\mathfrak{F}-\frac{\sigma'}{2r}-\frac{1}{2}\Bigg[\sigma''+\frac{\sigma'^{2}}{2}-\sigma'
r\mathfrak{F}\Bigg]&=&\frac{\aleph}{\mathfrak{F}}(P -
\frac{\chi}{2}\varrho^{2}-\frac{3}{2} \chi P^{2}),
\end{eqnarray}
with $\mathfrak{F}=\frac{\aleph}{1+\aleph r^{2}}$. The
non-conservation equation (\ref{4}) gives
\begin{equation}\label{17*}
\frac{dP}{dr}+{\sigma'}(\frac{P+\varrho}{2})+\ss^\star=0,
\end{equation}
where $\ss^\star$ determines the impact of $f(\Re, \mathcal{T}^{2})$
theory as
\begin{equation}\nonumber
\ss^\star=\Bigg[\sigma '\left(4
P\varrho+3P^2+\varrho^2\right)+(3PP'+\varrho\varrho')\Bigg]\frac{\chi}{2}.
\end{equation}

\section{Structure of Gravastar}

A thin shell of ultra-relativistic fluid covers the internal region
of gravastar, while the Schwarzschild metric describes the exterior
vacuum region. The following three domains provide a complete
picture of the gravstar structure.\\\\
(A) Interior domain $(\mathrm{J}_{1})$ $(0 \leq
r<\mathcal{K}_{1}=\mathcal{K})$ \quad$ \Longrightarrow\quad
\varrho=-P $.\\\\
(B) Intrinsic shell $(\mathrm{J}_{2})$ $(\mathcal{K}_{1}\leq r\leq
\mathcal{K}_{2})$
 \quad $\Longrightarrow \quad P=\varrho $.\\\\
(C) Exterior domain $(\mathrm{J}_{3})$
$(r>\mathcal{K}_{2}=\mathcal{K}+\epsilon)$ \quad
$\Longrightarrow\quad
\varrho=0=P$.\\\\
The small thickness of the intermediate shell is described by
$\epsilon$.

\subsection{The Internal Geometry}

We adopt the EoS, $P=\mathcal{W}\varrho$, where $\mathcal{W}$ is the
EoS parameter, as proposed by Mazur and Mottola \cite{13m}. The
inner region of gravastar structure is described by $P=-\varrho$
(for $\mathcal{W}=-1$). Consequently, a large amount of negative
pressure produced in core is applied on the intrinsic shell of the
spherical object and counterbalances the outward gravitational
force. Equation (\ref{17*}) alongwith dark energy EoS yields
$\varrho=\varrho_{c}$ ({\verb"constant"}) which gives
\begin{equation}\label{19}
P=-\varrho=-\varrho_{c}.
\end{equation}
This equation reads that matter variables (\emph{pressure as well as
energy density}) are constants throughout the interior region. The
other obtained solution from Eq.(\ref{17*}) is $(1+2{\chi}P)=0$
which gives $P=\frac{-1}{2{\chi}}$. This relation suggests that only
when the model parameter has a positive value, we obtain negative
pressure in the current setup which is necessarily required in the
inner core to prevent collapse. This negative pressure and
inward-acting force coming from shell counteract each other. In
order to accurately reflect this notion, our discussion specifically
considers the positive values of the model parameter. We determine
the radial metric component as
\begin{equation}\label{20}
e^{\varpi}=e^{(\mathcal{F}-4\mathcal{G}\varrho_{c}+C_{1})},
\end{equation}
where $\mathcal{F}=\frac{\aleph r^{2}}{2}$,
$\mathcal{G}=(\frac{r^{2}}{2}+\frac{a r^{4}}{4})(2\pi+\chi)$ and
$C_{1}$ denotes an integration constant.

The mass of celestial body for regular energy density is defined by
$M=\int4\pi\varrho r^{2}dr$, whereas in realm of
$f(\Re,\mathcal{T}^{2})$, the effective energy density leads to
\begin{equation}\label{22}
\mathbb{M}(\mathcal{K})=\frac{{\varrho_{c}(4\pi+\chi{\varrho_{c}})
}\mathcal{K}^{3}}{3}.
\end{equation}
Figure \textbf{1} shows the graphs of the gravitational mass
corresponding to regular density denoted by $M$ whereas $\mathbb{M}$
describes the gravitational mass corresponding to effective energy
density. It is obvious that thin-shell mass increases exponentially
in relation to thickness of the shell. This is the expected behavior
(for example, same outcomes were found for gravastars, employing
conformal motion in EMSG gravity \cite{zz}).
\begin{figure}\center
\epsfig{file=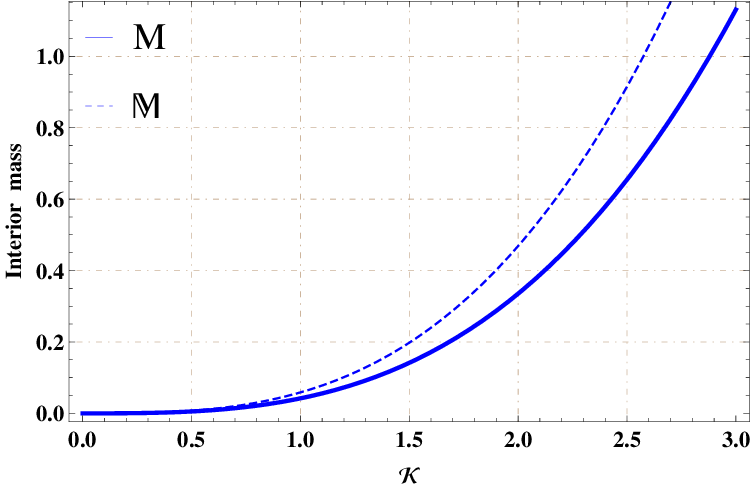,width=0.520\linewidth} \caption{Graphical
trend of gravitational mass corresponding to regular energy density
(Thick) and effective energy density (Dotted) for $\chi=2.5 $.}
\end{figure}

\subsection{The Intermediate Shell}

The correlation, $P=\varrho$, is followed by the intermediate shell
region (occupying stiff fluid). The stiff matter configuration
proposed by Zel'dovich \cite{38*} was used by researchers to have
some interesting results regarding cosmic and astrophysical issues
\cite{39}-\cite{43}. In present scenario, exact solution of density
is not possible. Thus, the interpolating function of density is
determined by incorporating stiff matter EoS whose graphical
behavior is observed in Figure \textbf{2}. The plot of density shows
the positive linear profile.
\begin{figure}\center
\epsfig{file=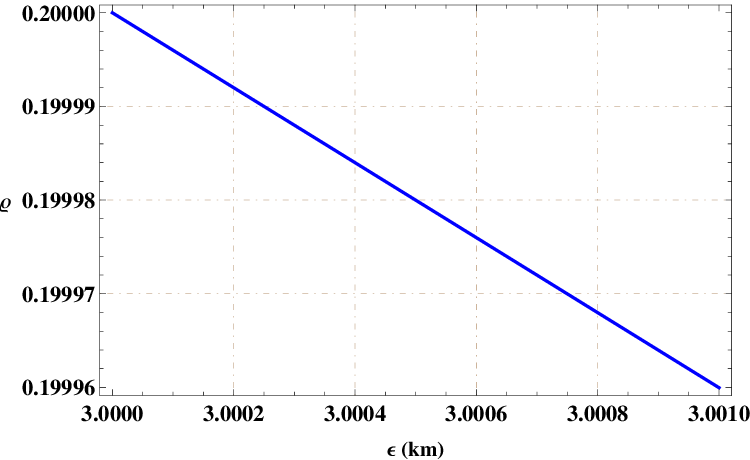,width=0.5\linewidth} \caption{Trend of energy
density against $\epsilon$ (shell thickness) for $\chi=2.5$.}
\end{figure}

\subsection{The External Region and Israel Matching Constraint}

To comprehend the unknown factors, it is essential to explore
certain physical constraints that are useful in investigating the
inner region of celestial bodies. By establishing a direct
correlation between the inner and outer regions of stellar objects,
we can effectively analyze the exact composition and behavior of
heavenly structures. In this pursuit, physicists often rely on
different spacetimes to examine the outer geometry of stellar
objects under various conditions. In the context of general
relativity (GR), the Schwarzschild spacetime is employed to study
non-charged objects, whereas the Reissner-Nordstrom metric is used
for charged objects. In recent study conducted by Kalita and
Mukhopaddhyay \cite{mu}, they investigated a modified asymptotically
flat vacuum solution, which is considered the best candidate for
external spacetime to facilitate smooth matching in modified
theories of gravity.

The modified spherically symmetric vacuum solution suitable for
representing the external region of stellar objects is given as
follows \setcounter{equation}{19}\begin{eqnarray}\label{14}
ds^{2}_{+}=\textsf{S}\mathrm{d}t^{2}-\frac{\mathfrak{P}}{\textsf{S}}
\mathrm{d}r^{2}-r^{2}d\theta^{2}-r^{2}\sin^{2}\theta d\phi^{2},
\end{eqnarray}
where
\begin{eqnarray}\nonumber
\textsf{S}&=&1-\frac{2}{r}-\frac{\textsf{B}(\textsf{B}-6)}{2r^2}+\frac{\textsf{B}^2(13\textsf{B}-66)}{20
r^3}-\frac{\textsf{B}^3(31\textsf{B}-156)}{48r^4}+\frac{3\textsf{B}^4(11\textsf{B}-57)}{56
r^5}
\\\nonumber
&-&\frac{\textsf{B}^5(67\textsf{B}-360)}{128r^6}+..., \quad
\mathfrak{P}=\frac{16r^4}{(\textsf{B}+2r)^4}.
\end{eqnarray}
Here $\textsf{B}$ is a real constant. It is mentioned here that two
boundaries complete the structure of gravastar, one connects the
inner region to the intermediate shell, whereas the other connects
the intermediate shell to the exterior spacetime. At these
boundaries, the continuity of metric potentials define physically
well behaved system. It is very crucial to determine the constraints
necessary for the smooth joining of both inner as well as outer
domains. Israel formalism plays vital role in attaining these
constraints which make possible the smooth joining of these domains.
The metric coefficients must be continuous over the hypersurface
($r=\mathcal{K}$), but the continuity between their differentials
may not hold.

The Lanczos equation is employed to compute the stress-energy tensor
as follows
\begin{equation}\label{25}
S_{\xi}^{\zeta}=(\delta_{\xi}^{\zeta}
\tau_{l}^{l}-\Im_{\xi}^{\zeta})\frac{1}{8\pi},
\end{equation}
where
$\Im_{\zeta\xi}=\mathcal{G}^{+}_{\zeta\xi}-\mathcal{G}^{-}_{\zeta\xi}$
exhibits that the extrinsic curvature is discontinuous. Here,  $+$
sign symbolizes the outer region of the gravastar whereas $-$ sign
corresponds to the inner region. The components of extrinsic
curvature at the hypersurface are described by
\begin{equation}\label{26}
{\mathcal{G}_{\zeta\xi}^{\pm}}= -n_{\gamma}^\pm
\left[\frac{\partial^2 x^{\gamma}}{\partial \varsigma^{\zeta}
\varsigma^{\xi}}+\Upsilon^{\gamma}_{\lambda\delta}\left(\frac{\partial
x^{\lambda}}{\partial\varsigma^{\zeta}}\right)\left(\frac{\partial
x^{\delta}}{\partial\varsigma^{\xi}}\right)\right],
\end{equation}
where thin-shell coordinates are represented by $\varsigma^{\xi}$
and the unit normal ($n_{\gamma}^{\pm}$) is defined as
\begin{equation}\label{27}
n_{\gamma}^{\pm}=\pm\left|g^{\lambda\delta}\frac{\partial
\Pi}{\partial x^{\lambda}}\frac{\partial \Pi}{\partial
x^{\delta}}\right|^{-\frac{1}{2}}\frac{\partial \Pi}{\partial
x^{\gamma}}, \quad n_{\gamma} n^{\gamma}=1.
\end{equation}
The isotropic EMT is given as
$S_{\xi}^{\zeta}=\text{diag}(\mathcal{Y},-\mathcal{Z},
-\mathcal{Z})$, where the surface energy density and surface
pressure are represented by $\mathcal{Y}$ and $\mathcal{Z}$,
respectively and are given by Lanczos equations as
\begin{eqnarray}\label{34}
&&\mathcal{Y}=-\frac{1}{4\pi\mathcal{K}}\Big[\sqrt{\textsf{S}}
-\sqrt{e^{(\mathcal{F}-4\mathcal{G}\varrho_{c})}}\Big],\\
&&\mathcal{Z}=\frac{1}{8\pi\mathcal{K}}\Big[\frac{\textsf{S'}}
{\sqrt{\textsf{S}}}-
\frac{e^{(\mathcal{F}-4\mathcal{G}\varrho_{c})}(1+\mathcal{K}(\mathcal{F'}-4\mathcal{G'}\varrho_{c}))}
{{\sqrt{e^{(\mathcal{F}-4\mathcal{G}{\varrho}_{0})}}}}\Big]\label{fff}.
\end{eqnarray}
The surface energy density is used to calculate mass of the
intermediate shell and is given as
\begin{equation}\label{36}
{m}_{\emph{shell}}=4\pi\mathcal{K}^{2}\mathcal{Y}=\mathcal{K}
\Big[\sqrt{e^{(\mathcal{F}-4\mathcal{G}\varrho_{c})}}
-\sqrt{\textsf{S}}\Big].
\end{equation}
Consequently, the total mass of the gravastar structure is
\begin{equation}\label{37}
\mathcal{M}=\Big[{e^{(\mathcal{F}-4\mathcal{G}\varrho_{c})}}\Big]^\frac{1}{2}{m}_{\textit{shell}}+\frac{(4\pi+\chi{\varrho_{c}})
{2{\varrho_{c}}{\mathcal{K}^{3}}}}{3}-\frac{{m}_{\textit{shell}}^{2}}{2\mathcal{K}}.
\end{equation}

\section{Some Physical Features Gravastar Intrinsic-Shell}

This section investigates significant gravastar features (EoS
parameter, length, energy, entropy as well as surface redshift
versus thin-shell thickness).

\subsection{Equation of State Parameter}

At $r=\mathcal{K}$, we employ Eqs.(\ref{34}) and (\ref{fff}) to
compute EoS parameter for the thin-shell region of gravastar
geometry as
\begin{equation}\label{38}
\mathcal{W}=\frac{\mathcal{Z}}{\mathcal{Y}}=
\frac{\frac{1}{8\pi\mathcal{K}}\Big[\frac{{1-\frac{\mathcal{M}}{\mathcal{K}}}}
{\sqrt{\frac{\mathcal{K}-{2\mathcal{M}}}{\mathcal{K}}}}-
\frac{1+\aleph \mathcal{K}^{2}}{({1+\aleph
\mathcal{K}^{2}})^{\frac{5}{2}}}\Big]}
{-\frac{1}{4\pi\mathcal{K}}\Big[\sqrt{1-\frac{2\mathcal{M}}{\mathcal{K}}}
-\sqrt{\frac{1}{1+\aleph \mathcal{K}^{2}}}\Big]}.
\end{equation}
The validity of various stellar models is well described by
different values of the EoS parameter. For instance, $\mathcal{W}=0$
indicates flat surfaces comprising of non-relativistic fluid,
$\mathcal{W}=\frac{-1}{3}$ is for curved surfaces, $\mathcal{W}=-1$
for dark energy, $\mathcal{W}<-1$ represents phantom energy and
$\mathcal{W}\geq-1$ describes non-phantom regime. In present
scenario, interior domain occupies dark energy so we take
$\mathcal{W}=-1$ showing significance of gravastar structure.

\subsection{Length of thin-Shell}

The distance from the internal region boundary to the intermediate
shell determines the length of the gravastar structure. It is
expressed by
\begin{equation}\label{n6}
{\textit{L}}=\int_{\mathcal{K}}^{\mathcal{K}+\epsilon}\sqrt{e^{\varpi}}dr.
\end{equation}
Differentiating this with respect to $r$, it follows that
\begin{equation}\label{n7}
{\textit{L}}'(r)=(1+\aleph
r^{2})^\frac{1}{2}|_{\mathcal{K}}^{\mathcal{K}+\epsilon}.
\end{equation}
The behavior of length is examined graphically in Figure \textbf{3}
which indicates that thin-shell length increases with shell
thickness.
\begin{figure}\center
\epsfig{file=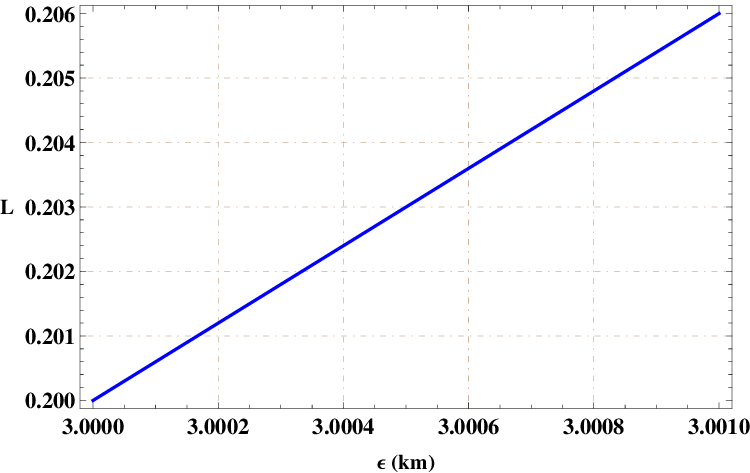,width=0.520\linewidth} \caption{Plot of proper
length versus thickness $\epsilon$ for $\chi=2.5$.}
\end{figure}

\subsection{Entropy of the Intermediate Shell}

The degree of disorder in a particular region of a structure is
referred to as entropy. The inner domain entropy is found to be zero
in literature \cite{13m}. The disorderness within intermediate shell
is described by
\begin{equation}\label{35}
S=4\pi\int_{\mathcal{K}}^{\mathcal{K}+\epsilon}\textbf{s(r)}\sqrt{e^{\varpi}}
r^{2} dr,
\end{equation}
where $\textbf{s(r)}$ is the entropy density. The plot of entropy
(Figure \textbf{4}) illustrates the continuous increasing profile of
randomness of the shell region. This indicates that the dark source
terms of $f(\Re,\mathcal{T}^2)$ gravity enhance the randomness of
the celestial object.
\begin{figure}\center
\epsfig{file=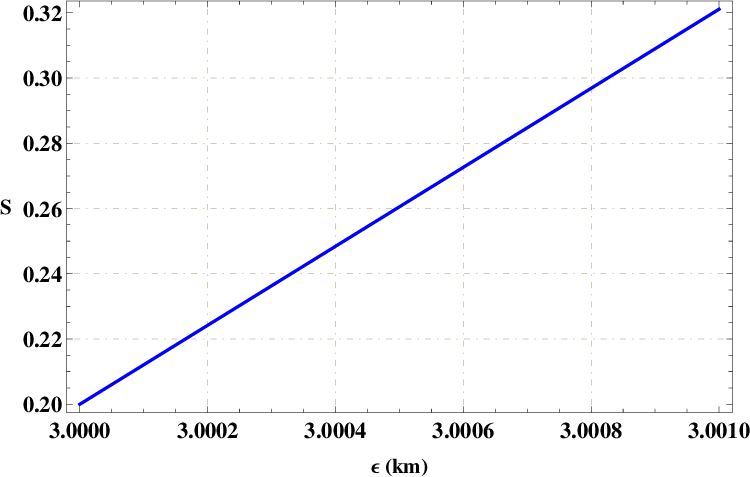,width=0.520\linewidth} \caption{Entropy
corresponding to thickness of intrinsic shell for $\chi=2.5$.}
\end{figure}

\subsection{Energy of Intrinsic-Shell}

The internal domain of the gravastar geometry occupies dark energy
providing anti-attractive force whereas energy of the shell region
is provided by
\begin{equation}\label{39}
{\varepsilon}=\int_{\mathcal{K}}^{\mathcal{K}+\epsilon}\frac{1}{2}\varrho
r^{2}dr.
\end{equation}
Figure \textbf{5} shows the increasing profile of the thin-shell
energy versus $\epsilon$.
\begin{figure}\center
\epsfig{file=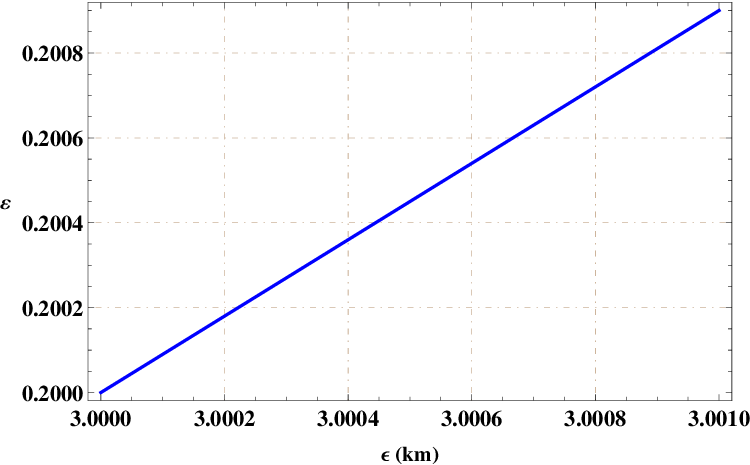,width=0.520\linewidth} \caption{Energy profile
against thickness of the shell for $\chi=2.5$.}
\end{figure}

\subsection{Surface Redshift of Intrinsic Shell}

The surface redshift is considered to be an important information
source for the stability and detection of gravastars in the
structural investigation of gravastars. The redshift parameter is
always less than 2 for isotropic matter distribution \cite{49}. It
is given as
\begin{eqnarray}\label{42}
{\textsl{z}}=\frac{1}{\sqrt{|g_{rr}|}}-1.
\end{eqnarray}
Figure \textbf{6} shows that the value of redshift parameter remains
in its required bound within the shell of gravastar geometry.
\begin{figure}\center
\epsfig{file=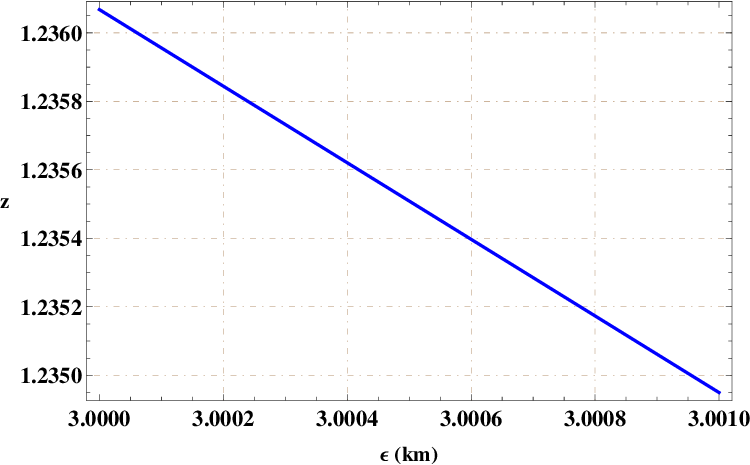,width=0.5\linewidth} \caption{Behavior of
surface redshift for $\chi=2.5$.}
\end{figure}

\section{Stability}

Here, we investigate system equilibrium to assess the stability of
the resulting solution through adiabatic index. Equation (\ref{17*})
involves two forces, i.e., hydrodynamical $\textsl{F}_{\mathbf{h}}$
and gravitational force $\textsl{F}_{\mathbf{g}}$ given by
\begin{eqnarray}\nonumber
\textsl{F}_{\mathbf{h}}&=&\frac{dP}{dr}+3\chi PP',\\\nonumber
\textsl{F}_{\mathbf{g}}&=&\sigma'(\frac{\varrho+P}{2})+\chi\left(\frac{3P^2+\varrho^2+4
P\varrho}{2}\right)\sigma'+\chi\varrho\varrho'.
\end{eqnarray}
The above forces describe equilibrium state by counterbalancing the
impact of each other. These forces include squared as well as
product matter components and combined effect of these forces turn
out to be zero. To discuss the stability of celestial bodies, we
utilize the adiabatic index. The pioneering proposal of
Chandrasekhar helps in checking the dynamical stability of celestial
body through small adiabatic perturbations. Chandrasekhar found that
the adiabatic index must not exceed the value 4/3 for the
relativistic system to be stable and is described as
\begin{equation}\\\nonumber
\Gamma =\frac{P+\varrho}{P}\frac{dP}{d\varrho}.
\end{equation}
For the inner domain with EoS $P=-\varrho$, we obtain $\Gamma=0$ and
for the thin-shell domain with EoS $P=\varrho$, it follows that
$\Gamma=2$. This shows that the internal sector is unstable but
intermediate shell sector is stable.

\section{Conclusions}

In this paper, we have computed solutions of gravastar geometry by
employing the Finch-Skea metric in the perspective of
${f}(\Re,\mathcal{T}^{2})$ gravity. The behavior of Finch-Skea
metric potential is regular and non-singular throughout gravastar
structure. In order to better understand different characteristics
of the gravastar structure, we have developed solutions for each
domain of the gravastar by using minimal coupling model
${f}(\Re,\mathcal{T}^{2})=\Re+\chi \mathcal{T}^{2}$. The metric
components as well as matter density in the internal region of the
gravastar are evaluated. The constant energy density is obtained
throughout the gravastar structure by employing $\varrho=-P$ in the
conservation equation. The exterior domain exerts pressure on
thin-shell and singularity formation is avoided by maintaining
balance between outward directed pressure and gravitational force
acting inward. Consequently, the singularity is prevented at the
core of the gravastar. We have observed that gravitational mass is
positive throughout the inner region but zero at the core.

We have also explored attributes of gravastar structure such as
matter density, EoS parameter, length, entropy, energy and surface
redshift against intrinsic shell thickness. The major findings are
summarized as follows.
\begin{itemize}
\item The finite positive behavior of density
is found versus shell thickness. As the thickness of a thin-shell
increases, so does the density (Figure \textbf{2}). Thus the shell
outer boundary is denser than its inner boundary.
\item
The gravastar length against intermediate shell shows an increasing
profile (Figure \textbf{3}). In $f(\Re,\mathcal{T}^{2})$, the value
of length is observed as $0.206$ which is less than that of
$2.6217008$ found in $f(\Re,\mathcal{T})$ theory \cite{gg} .
\item
The entropy of gravastar structure is linearly connected to
thickness of thin-shell. We have found higher entropy with
increasing thickness (Figure \textbf{4}). The shell region entropy
is $0.32$ for $\epsilon=3.0010km$ in the present scenario while its
value is found to be $2.1630$ in $f(\Re, \mathcal{T})$ gravity
\cite{gg}. It is notable that disorderness of the considered system
in realm of $f(\Re,\mathcal{T}^{2})$ theory is less than that of
$f(\Re, \mathcal{T})$ gravity.
\item
We have found linear behavior of the energy of gravastar thin-shell
versus shell thickness (Figure \textbf{5}). We have computed that
energy of the shell region is $0.2008$ for $\epsilon=0.30010km$
while in other scenario is $0.3538$ \cite{gg}. Thus, the shell
energy is decreased for this choice of EMSG model.
\item
The surface red shift provides the acceptable behavior as it lies in
the required limit.
\item
Finally, we have checked the hydrostatic equilibrium to analyze
stability of the system. We have also discussed the adiabatic index
to study the stability of gravastar. It is found that the stellar
model meets the specified range of stability criteria.
\end{itemize}

We would like to mention here that various studies on gravastar
structure in the realm of modified theories of gravity have been
carried out \cite{c1}-\cite{c3}. The discussed attributes, i.e.,
length, entropy and energy present proportional behavior. In
comparison to our obtained outcomes, we have found consistent
behavior with those given in the literature \cite{c1}-\cite{c3}.
These features demonstrate physical viability as well as stability
of the considered structure in the context of
${f}(\Re,\mathcal{T}^{2})$ theory. Thus the construction of a
gravastar-like compact object with Finch-Skea ansatz is realistic in
this modified theory.\\\\
\textbf{Data Availability Statement:} This manuscript does not
contain any new data.


\begin{thebibliography}{37}

\bibitem{sc} Pietrobon, D., Balbi, A. and Marinucci, D.: Phys. Rev. D \textbf{74}(2006)043524;
Astieer, P. et al.: Astron. Astrophys.
\textbf{447}(2016)31.

\bibitem{7a} Katirci, N. and Kavuk, M.: Eur. Phys. J. Plus \textbf{129}(2014)163.

\bibitem{36} Roshan, M. and Shojai, F.: Phys. Rev. D \textbf{94}(2016)044002.

\bibitem{37} Board, C.V.R. and Barrow, J.D.: Phys. Rev. D \textbf{96}(2017)123517.

\bibitem{38} Nari, N. and Roshan, M.: Phys. Rev. D \textbf{98}(2018)024031.

\bibitem{17} Moraes, P.H.R.S. and Sahoo, P.K.: Phys. Rev. D \textbf{97}(2018)2.

\bibitem{18} Akarsu, O. et al.:  Phys. Rev. D \textbf{97}(2018)124017.

\bibitem{41} Bahamonde, S., Marciu, M. and Rudra, P.:  Phys. Rev. D \textbf{100}(2019)083511.

\bibitem{21} Barbar, A.H., Awad, A.M. and AlFiky, M.T.: Phys. Rev. D \textbf{101}(2020)044058.

\bibitem{22} Ranjit, C. Rudra, P. and Kundu, S.: Ann. Phys. \textbf{428}(2021)168432.

\bibitem{23} Singh, K.N. et al:  Phys. Dark Universe \textbf{31}(2021)100774.

\bibitem{42} Sharif, M. and Gul, M.Z.: Phys. Scr. \textbf{96}(2020)025002; Int. J. Mod. Phys. A
\textbf{36}(2021)2150004; Adv. Astron. \textbf{2021}(2021)6663502;
Eur. Phys. J. Plus \textbf{136}(2021)503; Chin. J. Phys.
\textbf{71}(2021)365; Universe \textbf{07}(2021)154; Phys. Scr.
\textbf{96}(2021)105001.

\bibitem{44} Sharif, M. and Naz, S.: Universe \textbf{ 8}(2022)142;  Eur. Phys. J. Plus \textbf{137}(2022)421;
Int. J. Mod. Phys. D \textbf{31}(2022)2240008; Mod. Phys. Lett. A
\textbf{37}(2022)2250065; ibid. 2250125.

\bibitem{44*} Sharif, M. and Iltaf, S.: Phys. Scr. \textbf{97}(2022)075002; Chin. J. Phys. \textbf{79}(2022)173.

\bibitem{44**} Sharif, M. and Anjum, A.: Eur. Phys. J. Plus \textbf{137}(2022)602; Gen. Relativ. Gravit
\textbf{54}(2022)111.

\bibitem{13m} Mazur, P. and Mottola, E.: Proc. Natl. Acad. Sci. \textbf{101}(2004)9545.

\bibitem{zz} Sharif, M. and Saeed, M.: Chin. J. Phys. \textbf{77}(2022)583.

\bibitem{d1} Sakai, N. et al.: Phys. Rev. D \textbf{90}(2014)104013.

\bibitem{d2} Kubo, T. and Sakai, N.: Phys. Rev. D \textbf{93}(2016)084051.

\bibitem{d3} Cardoso, V. et al.: Phys. Rev. Lett. \textbf{116}(2016)171101; ibid. \textbf{117}(2016)089902 .

\bibitem{d5} Akiyama, K. et al.: Astrophys. J. Lett. \textbf{875}(2019)1.

\bibitem{5} Visser, M. and Wiltshire, D.L.: Class. Quantum Grav. \textbf{21}(2004)1135.

\bibitem{6} Carter, B.M.N.: Class. Quantum Grav. \textbf{22}(2005)4551.

\bibitem{7} Bili\'{c}, N., Tupper, G.B. and Viollier, R.D.: J. Cosmol. Astropart. Phys. \textbf{02}(2006)013.

\bibitem{34} Ghosh, S. et al.: Res. Phys. \textbf{14}(2019)102473.

\bibitem{36*} Ghosh, S. et al.: Ann. Phys. \textbf{411}(2019)167968.

\bibitem{29} Das, A. et al.: Phys. Rev. D \textbf{95}(2017)124011.

\bibitem{29tt} Shamir, M. and Ahmad, M.: Phys. Rev. D \textbf{97}(2018)104031.

\bibitem{28} Sharif, M. and Waseem, A.: Astrophys. Space Sci. \textbf{364}(2019)189.

\bibitem{35} Sharif, M. and Waseem, A.: Int. J. Mod. Phys. D \textbf{28}(2019)1950033; ibid. 2040005.

\bibitem{702} Cruz-Dombriz, A. and Saez-Gomez, D.: Entropy \textbf{14}(2012)1717.

\bibitem{708} Staykov, K.V.: J. Cosmol. Astropar. Phys. \textbf{10}(2014)006.

\bibitem{711} Yazadjiev, S.S. et al.: J. Cosmol. Astropar. Phys. \textbf{06}(2014)003.

\bibitem{710} Yazadjiev, S.S., Doneva, D.D. and Kokkotas, K.D.: Phys. Rev. D
\textbf{91}(2015)084018; ibid. \textbf{92}(2015)064015; ibid.
\textbf{96}(2017)064002; Eur. Phys. J. C \textbf{78}(2018)818.

\bibitem{704} Capozziello, S. et al.: Phys. Rev. D \textbf{93}(2016)023501.

\bibitem{705} Doneva, D.D. et al.: Phys. Rev. D \textbf{98}(2018)104039.

\bibitem{701} Olmo, G.J., Rubiera-Garcia, D. and Wojnar, A.: Phys. Rept.
 \textbf{876}(2020)1.

\bibitem{703} Feola, P. et al.: Phys. Rev. D \textbf{101}(2020)044037.

\bibitem{712} Odintsov, S.D. and Oikonomou, V.K.: Phys. Rev. D \textbf{107}(2023)104039.

\bibitem{714} Oikonomou, V.K.: Class. Quantum Grav. \textbf{40}(2023)085005;
Monthly Notices Royal Astronom. Soc. \textbf{520}(2023)2934.

\bibitem{21} Dourah, H.L. and Ray, R.: Class. Quantum Grav. \textbf{4}(1987)1691.

\bibitem{22} Finch, M.R. and Skea, J.E.F.: Class. Quantum Grav. \textbf{6}(1989)467.

\bibitem{22a} Kalam, M. et al.: Int. J. Theor. Phys. \textbf{52}(2013)3319.

\bibitem{22b} Banerjee, A. et al.: Gen. Relativ. Gravit. \textbf{45}(2013)717.

\bibitem{22c} Hansraj, S. et al.: Int. J. Mod. Phys. D \textbf{15}(2006)1311.

\bibitem{22d} Hansraj, S. and Maharaj, S.D.: Int. J. Mod. Phys. D
\textbf{15}(2006)1311.

\bibitem{22e} Sharma, R., Das, S. and Thirukkanesh, S.: Astrophys.
Space Sci. \textbf{362}(2017)12.

\bibitem{22f} Paul, B.C. and Dey, S.: Astrophys. Space Sci. \textbf{363}(2018)220.

\bibitem{22g} Bhar, P.: Astrophys. Space Sci. \textbf{359}(2015)41.

\bibitem{ab} Akarsu, O. Katirci, N., and Kumar, S.: Phys. Rev. D \textbf{97}(2018)024011.

\bibitem{7b} Finch, M.R. and Skea, J.E.F.: Class. Quantum Grav. \textbf{4}(1989)467.

\bibitem{38*} Zel'dovich, Y.B.: Mon. Not. R. Astron. Soc. \textbf{160}(1972)1.

\bibitem{39} Carr, B.J.: Astrophys. J. \textbf{201}(1975)1.

\bibitem{40} Wesson, P.S.: J. Math. Phys. \textbf{19}(1978)2283.

\bibitem{41*} Madsen, M.S. et al.: Phys. Rev. D \textbf{46}(1992)1399.

\bibitem{42*} Braje, T.M. and Romani, R.W.: Astrophys. J. \textbf{580}(2002)1043.

\bibitem{43} Linares, L.P., Malheiro, M. and Ray, S.: Int. J. Mod. Phys. D \textbf{13}(2004)1355.

\bibitem{mu} Kalita, S. and Mukhopadhyay, B.: Eur. Phys. J. C \textbf{79}(2019)877.

\bibitem{49} Buchdahl, H.A.: Phys. Rev. \textbf{116}(1959)1027;
Barraco, D.E. and Hamity, V.H.: Phys. Rev. D
\textbf{65}(2002)124028.

\bibitem{gg} Majeeda, K., Abbas, G. and Siddiqa, A.: New Astron. \textbf{95}(2022)101802.

\bibitem{c1} Bhar, P.: Astrophys. Space Sci. \textbf{354}(2014)457;
Int. J. Geom. Methods Mod. Phys. \textbf{18}(2021)2150112.

\bibitem{c5}  Das, A. et al.: Phys. Rev. D \textbf{95}(2017)124011;
Das, A. et al.: Nucl. Phys. B \textbf{954}(2020)114986.

\bibitem{c3} Bhar, P and Rej, P.: Eur. Phys. J. C \textbf{81}(2021)763.

\end{thebibliography}
\end{document}